\documentclass[epsfig]{aa}
\usepackage{graphics}

\def\Teff{$T_{\rm eff}$}
\def\cc{\element[][12]{C}/\element[][13]{C}}
\def\coco{\element[][12]{CO}/\element[][13]{CO}}
\def\cod{\element[][12]{CO}}
\def\cot{\element[][13]{CO}}

\begin{document}

\thesaurus{06(          
              08.16.4;  
              08.03.1;  
              08.06.3;  
              08.12.1;  
              08.15.1)  
           }

\title{Optical and near-infrared spectrophotometric properties of Long 
Period Variables and other luminous red stars}
 
\author{R. Alvarez\inst{1}  
   \and A. Lan\c{c}on\inst{2}
   \and B. Plez\inst{3}
   \and P.R. Wood\inst{4}
}
\institute{IAA, 
           Universit\'e Libre de Bruxelles, 
           C.P.\,226, Bvd du Triomphe,
           B--1050 Bruxelles, Belgium \\
           {\tt ralvarez@astro.ulb.ac.be}
\and
           Observatoire de Strasbourg,
           UMR 7550,
           11 Rue de l'Universit\'e, 
           F--67000 Strasbourg, France \\
           {\tt lancon@astro.u-strasbg.fr}
\and
           GRAAL,
           Universit\'e Montpellier II,
           cc072,
           F--34095 Montpellier cedex 05, France \\
           {\tt plez@graal.univ-montp2.fr}
\and	
	   RSAA (ANU), Mount Stromlo Observatory,
	   Private Bag, Weston Creek, 
           ACT 2611, Australia \\
           {\tt wood@mso.anu.edu.au}
}

\offprints{R. Alvarez}
\mail{ralvarez@astro.ulb.ac.be}

\date{Received / Accepted}

\titlerunning{Spectrophotometric properties of LPVs and other red stars}
\authorrunning{R.\ Alvarez et al.}

\maketitle

\begin{abstract}
Based on a new and large sample of optical and near-infrared spectra 
obtained at the Mount Stromlo and Siding Spring Observatories (Lan\c{c}on 
\& Wood 1998; Lan\c{c}on \& Wood, in preparation), spectrophotometric 
properties of cool oxygen- and carbon-rich Long Period Variables and 
supergiants are presented. 
Temperatures of oxygen-rich stars are assigned by comparison with synthetic 
spectra computed from up-to-date oxygen-rich model atmosphere grids. 
The existence of reliable optical and near-infrared temperature indicators 
is investigated. 
A narrow relation between the bolometric correction BC$_{\rm I}$ and the 
broad-band colour I$-$J is obtained for oxygen-rich cool stars.
The ability of specific near-infrared indices to separate luminosity 
classes, atmospheric chemistry or variability subtypes is discussed. 
Some comments are also given on extinction effects, water band strengths 
in Long Period Variables and the evaluation of \coco\ ratio in red giants.
\keywords{Stars: AGB and post-AGB -- Stars: carbon -- 
Stars: fundamental parameters -- Stars: late-type -- Stars: oscillations}
\end{abstract}

\section{Introduction}

The near-infrared (NIR) radiation of our and other galaxies is essentially
provided by luminous red stars: red supergiants dominate in starburst 
regions, Asymptotic Giant Branch stars (AGBs), among which the O-rich 
and C-rich Long Period Variables (LPVs) of various variability types, take 
over in stellar populations with ages between $10^8$ and $10^9$ years, and 
First Giant Branch stars dominate in older regions (Bruzual \& Charlot 1993;
Lan\c{c}on 1999).
In view of the recent developments in NIR observational technology, tools 
have become necessary that would allow us to connect NIR spectrophotometric 
signatures with the fundamental stellar properties provided by theoretical 
stellar evolution tracks. Only then can integrated NIR emission 
properties or star counts be interpreted in terms of galaxy history.
These connection tools are also needed from a purely stellar perspective. 
The evolution, structure and atmospheres of the evolved red stars are among 
the standing problems in stellar astrophysics and valuable constraints on 
the models are to be expected from population studies. The choice of 
suitable filter passbands, able to separate stellar subtypes, is a 
prerequisite for efficient surveys.

In this paper, the emphasis is set on LPVs, which constitute the most 
luminous and coolest but also the most complex population of red giants 
in regions where more massive red supergiants have died. Static red 
giant model atmospheres are progressing rapidly due to recent improvements 
in molecular opacity data (Plez 1999a, 1999b).
Variable star models are still far from satisfactory because of the 
interplay between pulsation, radiative transfer and chemistry (see H\"ofner 
1999 for a review).
Empirical spectroscopic or spectrophotometric studies adequate for our
purpose have remained scarce because they are time consuming and because a 
significant wavelength coverage requires synchronised observations: LPV 
periods are typically one year long and cycle to cycle variations common. 
In addition, since good infrared instrumentation is usually found on 
relatively large telescopes that tend to be preferentially directed towards 
faint sources instead of bright stars, most previous photometric studies 
of LPVs have been optical (Kerschbaum \& Hron 1996; Alvarez \& Mennessier 
1997; Wing et al.\ 1998). Based on a new, rich sample of complete 
optical+NIR spectra of red stars, we are able to extend these studies. 

The properties of the data are summarised in Sect.~2. In Sect.~3, we discuss 
spectrophotometric temperature indicators at optical and NIR wavelengths.
The physical significance and absolute calibration of the empirical scales,
in particular for variable stars, will be discussed in more depth elsewhere. 
The subsequent sections discuss the ability of commonly used and newly 
defined NIR narrow band indices to separate luminosity, surface chemistry or 
variability subtypes. The results are summarised in Sect.~10.

\section{The sample}

The reader is refered to Lan\c{c}on \& Wood (1998) and Lan\c{c}on \& Wood, 
in preparation, for a thorough description of the sample. Around one hundred 
spectra of cool stars ranging from 5100~\AA\ to 2.5~$\mu$m have been obtained 
at the Mount Stromlo and Siding Spring Observatories, using the three 
cross-dispersed grisms of CASPIR\footnote{Cryogenic Array Spectrometer and 
Imager, McGregor 1994} on the 2.3~m Telescope (near-infrared spectra with a 
resolution R of 1100) and a low resolution grating spectrograph on the 
74~inch Telescope (optical spectra, R $\simeq 30$). The sample mainly 
contains instantaneous spectra of O-rich and C-rich local LPVs\footnote{All 
variability types used in this paper are from the {\it General Catalogue of 
Variable Stars} (Kholopov et al.\ 1988)}, but also includes local 
non-variable giants and a few luminous LPVs from the Magellanic Clouds and 
the Galactic Bulge. Additional spectra restricted to either the optical 
(5100--9900~\AA) or the NIR region (9750--24900~\AA) are available for some 
stars and are included in the analysis whenever possible.

\section{Temperature indicators}

\subsection{Temperature definitions}

Only little is known about effective temperatures of LPVs. In principle, 
the effective temperature relates the bolometric luminosity of a star to 
its effective radius and is thus accessible through surface brightness 
measurements, i.e.\ apparent luminosities combined with angular diameters 
(e.g.\ Perrin et al.\ 1998). However, due to their very extended 
atmospheres, LPVs such as Miras cannot easily be assigned a unique radius: 
observational determinations are wavelength dependent and depend on the 
adopted limb-darkening law (Labeyrie et al.\ 1977; Baschek et al.\ 1991; 
Hofmann et al.\ 1998). The use of model atmospheres is inevitable not only 
in providing these laws in the various observational filter passbands, but 
also to relate the resulting ``monochromatic'' radii to the effective 
radius (usually based on Rosseland mean optical depths) that is used in 
the effective temperature definition of stellar models.
Large uncertainties and inconsistencies between authors occur as a result 
of e.g.\ incomplete treatment of molecular opacities in the models, of 
different assumptions on the stellar pulsation mode or on whether or not the 
phase in the pulsation cycle is taken into account (Scholz \& Takeda 1987; 
Haniff et al.\ 1995; van Belle et al.\ 1996; Bessell et al.\ 1996).

Photometric indices, i.e.\ colours and molecular band strength measurements, 
provide alternative effective temperature estimates (Ram\'{\i}rez et al.\ 
1997; Alvarez \& Mennessier 1997).  Optical indices define the spectral 
types, which are calibrated against the effective temperature either on the 
basis of angular diameter measurements (Ridgway et al.\ 1980; Perrin et al.\ 
1998) or by direct comparison with synthetic spectra (Fluks et al. 1994). 
The calibrations inherit the model dependence already mentioned. In addition 
they generally assume a unique relation between spectral type and
effective temperature, independent of variability type or time.
This assumption oversimplifies the real situation: Alvarez \& Plez (1998) 
have shown that TiO and VO bands, two temperature sensitive indicators, do 
not vary in phase along the Mira luminosity cycle, and the data used here 
contains pairs of spectra with insignificant differences below 1~$\mu$m but 
very different NIR H$_2$O vapour bands. 
As expected from the complex stratification of dynamical atmosphere models 
(Bowen 1988; Fleischer et al.\ 1992; H\"ofner et al.\ 1998) molecular indices 
contain local information and are sensitive each in its own way to the 
details of the pulsation history as well as to the global property called 
effective temperature.

In view of the above difficulties, this paper avoids direct comparison 
with dynamical model atmospheres. A temperature refered to as the 
{\em ``static'' effective temperature} (and noted \Teff\ throughout
the paper for simplicity)\footnote{``Effective temperature", {\em in 
extenso}, is used when the theoretical quantity is meant.} is nevertheless 
tentatively assigned to each spectrum, based on the fact that the optical 
part of the static and variable giant spectra in our sample can indeed be 
reasonably well fitted with static giant models. The resulting absolute 
\Teff\ scales may be biased estimates of the real effective temperatures, 
but a better understanding of pulsating models is required before these 
biases can be evaluated and their dependence on the pulsation properties 
accounted for. The static \Teff\ scale is used to calibrate photometric 
temperature scales and all further discussion then relies on purely 
empirical findings.

\subsection{Static effective temperature assignment}

The synthetic spectra used for the static \Teff\ assignment are computed 
from a large grid of MARCS models (Plez et al.\ 1992; Plez 1992; Plez,
private communication) representing red giants with solar composition and 
different values of atmospheric extension, surface gravities $\log g$ 
ranging from $-$0.5 to 3.5, and of effective temperatures from 2400 to 
4750~K. The spectral synthesis includes the updated opacity data for TiO 
(Plez 1998), VO and H$_2$O as detailed in Bessell et al.\ (1998) and 
Alvarez \& Plez (1998). The latest models are preliminary computations of 
a large new and updated grid now being computed by the Uppsala group.

The static \Teff\ associated with a spectrum is defined here as the 
effective temperature of the MARCS model that provides the minimum $\chi^2$ 
fit to the 5100--9950~\AA\ data, $\log g$ and atmospheric extension being 
allowed to vary freely. 

\begin{figure}
  \resizebox{\hsize}{!}{\includegraphics{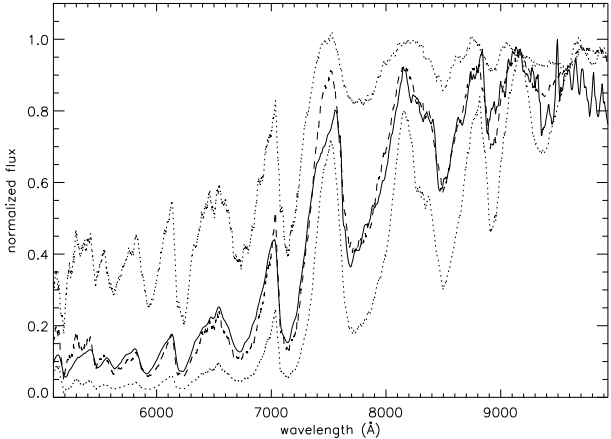}}
  \caption[]{Observed spectrum of S\,Phe, a M6 SRb variable (full line)
   and synthetic spectrum with \Teff=3400~K, $\log g$=1.0 and $\cal M$=1.0
   $\cal M_{\odot}$ (dashed line). The dotted lines are two synthetic 
   spectra with \Teff=3200~K and 3600~K}
  \label{cs}
\end{figure}

We have determined this \Teff\ for 131 spectra of oxygen-rich giants, 
among which some are Mira, Semi-Regular (SRa and SRb types) or Irregular 
variables, observed at various phases. The sensitivity of the spectra to 
\Teff\ combined with the step of the model grid allows us to assign \Teff\ 
values to $\pm$100~K (cf.\ Fig.~\ref{cs}). 
The corresponding acceptable $\log g$ values lie between 0.0 and 1.5, a 
range over which spectral variations are much smaller than those produced 
by a 100~K change in \Teff . 

\subsection{Photometric temperature indicators}

The following sections discuss narrow band and broad band colours derived 
from the observed spectra. We use filters similar to those of Wing (1967) 
and White \& Wing (1978), adopting the passbands defined by Bessell et 
al.\ (1989a); we complete the set with filters from Fluks et al.\ (1994) 
and additional indices designed to measure OH, H$_2$O and CO band strengths.
The VRIJHK colours are computed with the passbands of Bessell \& Brett 
(1988) and Bessell (1990). The zero point calibrations are based on a 
spectrum of Vega (Castelli \& Kurucz 1994) as used by Bessell et al.\ 
(1998): the colours of Vega are set equal to zero, except for the Wing 
colours for which we used Vega's observed colours (Wing, private 
communication).

\subsubsection{Optical temperature indicators}

Most optical colours appear to be quite good temperature indicators
(e.g.\ 710$-$888, 847$-$839, 710$-$745, 888$-$883 as used by Wing, 
or $S_{1/2, Sp}$, $S_{2/3, Sp}$ as in Fluks et al.\ 1994).

\begin{figure}
  \resizebox{\hsize}{!}{\includegraphics{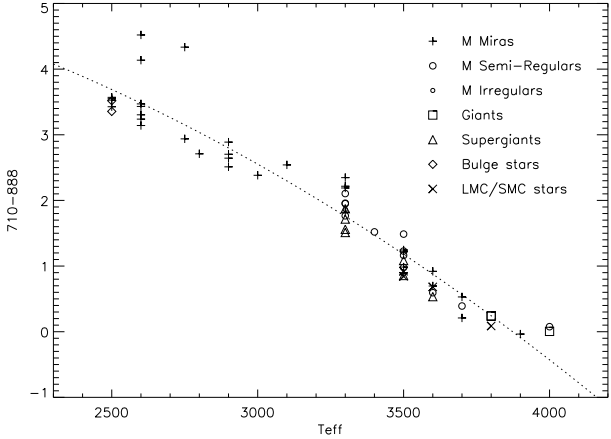}}
  \caption[]{The 710$-$888 index as a function of the effective
  temperature derived from the fit of MARCS models to the optical spectrum
  (plus signs: M Miras; empty circles: M Semi-Regulars; small empty circles: 
  M Irregulars; squares: K and M giants; triangles: supergiants; diamonds: 
  bulge stars; crosses (X): LMC/SMC stars}
  \label{to}
\end{figure}

\begin{figure}
  \resizebox{\hsize}{!}{\includegraphics{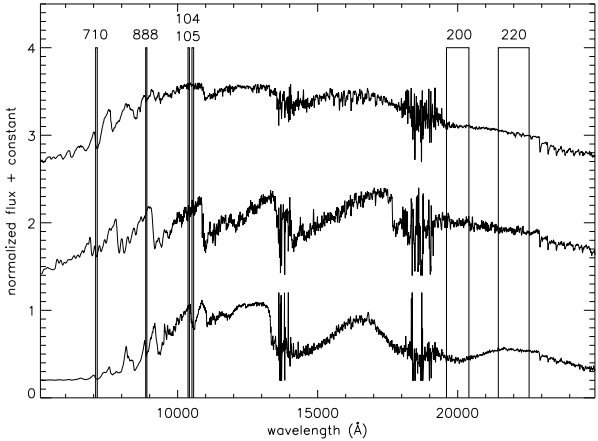}}
  \caption[]{Narrow-band filters used in Sect.~3. The spectra are
  from bottom to top: R\,Phe, a O-rich Mira; RU\,Pup, a C-rich 
  Semi-Regular and HD\,101712, a supergiant. Regions around 1.4 and
  1.9~$\mu$m are associated with extreme telluric water absorption}
  \label{vf1}
\end{figure}

Figure~\ref{to} shows the 710$-$888 index as a function of the \Teff\ 
assigned previously. This index has been chosen to be our
temperature indicator subsequently as its scale of variation is large and 
the scatter is small. It measures the ratio between the bandheads of the 
TiO $\gamma$ $\Delta$v=0 and the TiO $\delta$ $\Delta$v=0 ro-vibrational 
transitions (Brett 1990). The 710 and 888 filters are displayed on 
Fig.~\ref{vf1}.
A least-square polynomial fit gives:
\begin{equation}
 (710-888) = 5.8 + 3.2 \, 10^{-4} \times T_\mathrm{eff}
             - 4.7 \, 10^{-7} \times T_\mathrm{eff}^2
 \label{fitT}
\end{equation}
The standard deviation is 0.3~mag in the range 2500~K $\leq$ \Teff\
$\leq$ 3900~K. 

It is worth noticing that non-variable and variable stars on one hand, 
local, Bulge and LMC/SMC stars on the other hand, all seem to follow the 
same relation. This may be surprising at first thought. In fact, it shows 
that the relation between the 710$-$888 index and the whole optical spectrum 
used in our $\chi^2$ adjustment doesn't depend strongly on metallicity or 
variability type. The optical parts of the spectra all tend to lie on a 
tight sequence (which makes the fits with static model spectra possible).
It doesn't imply, however, that the effective temperature scale along this 
sequence is independent of the object type or metallicity: our \Teff\ 
assignment based on the optical spectrum is liable to be a good estimate of 
the real effective temperature of static stars, but could be systematically 
different from the effective temperature in variable objects.

Only two Bulge stars of our sample were observed in the optical region.
More observations are required to make firm statements on their specific 
static \Teff\ versus (710$-$888) relation.

The outlying points in Fig.~\ref{to} at 710$-$888 $\geq$ 4 are three 
observations of SV\,Lib, a M8 Mira star. These values of the TiO index 
would lead to particularly low temperatures when extrapolating the 
\Teff--(710$-$888) relation without caution. However, the global energy 
distribution of the star and the discussion in the following section 
argue against such an interpretation. These points were not taken into 
account when deriving Eq.~(\ref{fitT}).

\subsubsection{Near-infrared temperature indicators}

For extragalactic study purposes, near-infrared temperature indicators 
are of great interest. Indeed, the contribution of AGB stars to the 
emission of a stellar population is largest in this spectral range, and the 
population synthesis predictions are thus particularly sensitive to 
temperature--colour--spectral feature relations (Lan\c{c}on 1999).

\begin{figure}
  \resizebox{\hsize}{!}{\includegraphics{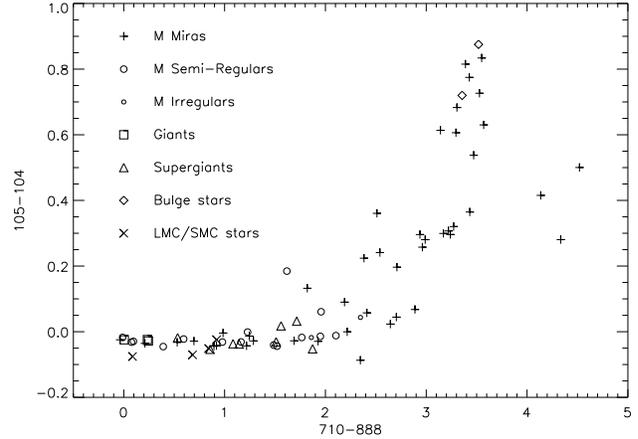}}
  \caption[]{The 105$-$104 index versus the \Teff\ indicator
   710$-$888}
  \label{ti}
\end{figure}

We have applied various near-infrared filters to the observed spectra. In 
Fig.~\ref{ti}, the Wing VO index 105$-$104 is plotted against our optical 
\Teff\ indicator 710$-$888. There is a good correlation between the two 
colours, the VO index becoming a useful \Teff\ index below 3300~K, as 
already found by Bessell et al. (1989a) from synthetic spectra. 
The 104 and 105 filters are shown in Fig.~\ref{vf1}.
The scatter of Fig.~\ref{ti} might be due to atmospheric extension,
surface gravity or metal abundance (Bessell et al.\ 1989a) but the main 
source of scatter is certainly variability, Miras executing loops in 
such TiO/VO diagrams (Bessell et al.\ 1996; Alvarez \& Plez 1998). 

The three outlying points (710$-$888$ \geq $4) are, again, the observations 
of SV\,Lib. This M8 Mira star has strong TiO absorption at 710~nm but 
normal VO bands (see also Lan\c{c}on \& Wood 1998). The observations were 
taken at different dates: the effect cannot be associated with errors in the 
acquisition or the reduction of the data. This star deserves further 
scrutiny and in particular a more complete phase coverage to clarify its 
pecularities.

\begin{figure*}
  \resizebox{\hsize}{!}{\rotatebox{90}{\includegraphics{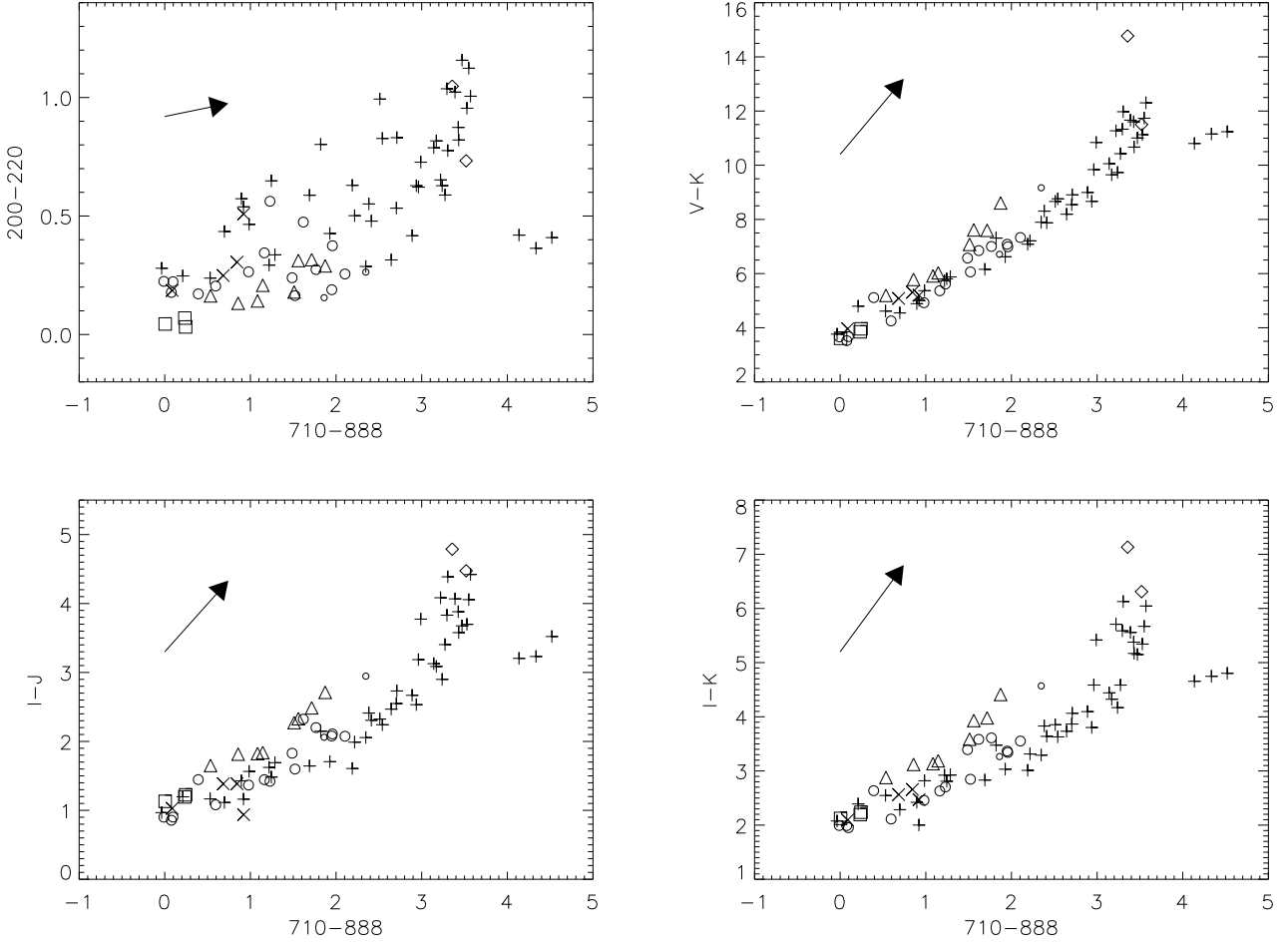}}}
  \caption[]{The 200$-$220, V$-$K, I$-$J and I$-$K indices versus 
  the temperature indicator 710$-$888. The arrows give the extinction 
  vectors for an $E_{B-V}$ equal to 1 (see Sect.~5). Same symbols as 
  Fig.~\ref{ti}}
  \label{ti+}
\end{figure*}

It is difficult to define near-infrared colours more convenient than
105$-$104 that might be used as \Teff\ indicators. 
The CO (2.3~$\mu$m) equivalent width suggested by Ram\'{\i}rez et al.\ 
(1997) appears inadequate: the correlation with \Teff\ is only 
marginally significant in our LPV sample. The large dispersion is mainly 
due to variations of the order of 10~\AA\ in the CO equivalent width of 
individual stars, which are not in phase with the \Teff\ variations. 
Differences in metallicity and C/O ratios are likely to contribute to the 
dispersion.
Figure~\ref{ti+} shows the indices most tightly correlated with 710$-$888. 
The 200$-$220 index which is a measure of the strength of H$_2$O at 2.0 
$\mu$m relative to the continuum point at 2.2 $\mu$m reveals a more 
pronounced scatter than the VO index when plotted against the
\Teff\ indicator 710$-$888. The 200 and 220 filters are displayed on 
Fig.~\ref{vf1}.

\begin{figure}
  \resizebox{\hsize}{!}{\includegraphics{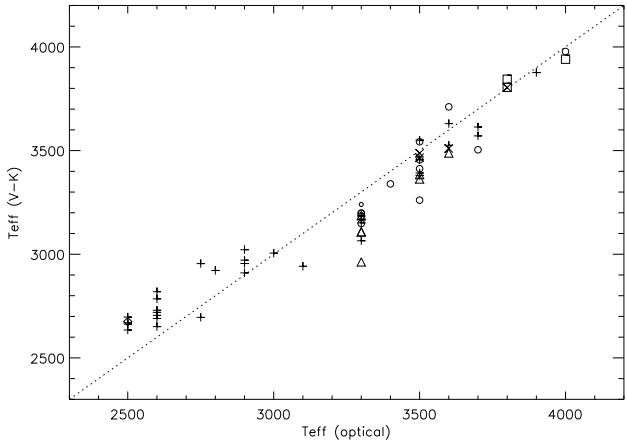}}
  \caption[]{The effective temperature obtained from V$-$K and the relation 
  of Bessell et al.\ (1998) against our static \Teff\ assignment derived
  from the fit of MARCS models to the optical spectrum (see Sect.~3.2).
  The dotted line is the one-to-one line. Same symbols as Fig.~\ref{ti}}
  \label{ct}
\end{figure}

The broad-band colour V$-$K is not a pure near-infrared index. It is a 
very good effective temperature index for static cool stars, not so much 
affected by metallicity effects according to Bessell et al. (1998), and we 
find it very well correlated with the 710$-$888 index. V$-$K values have 
been computed after extrapolating each observed spectrum (which begins at 
5100~\AA\ and ends at 24900~\AA) on the blue and on the red sides by the 
synthetic spectrum which gave the best $\chi^2$ fit. The correction is 
always less than 0.15~mag (mean: 0.09~mag).
Bessell et al.\ (1998) have determined polynomial fits to empirical 
effective temperature versus V$-$K relations for giants. Figure~\ref{ct} 
shows the comparison between the effective temperature obtained using their 
relation (their Table~7, data sets 'abcd') and the static \Teff\ we 
determined in Sect.~3.2: the agreement is good.
The broad-band colours I$-$J and I$-$K are almost as good \Teff\ indicators 
as V$-$K (Fig.~\ref{ti+}). A systematic difference between the supergiants 
(triangles) and the other stars is seen: their I$-$J and I$-$K tend to 
be larger at given 710$-$888.

\section{Bolometric corrections}

\begin{figure}
  \resizebox{\hsize}{!}{\includegraphics{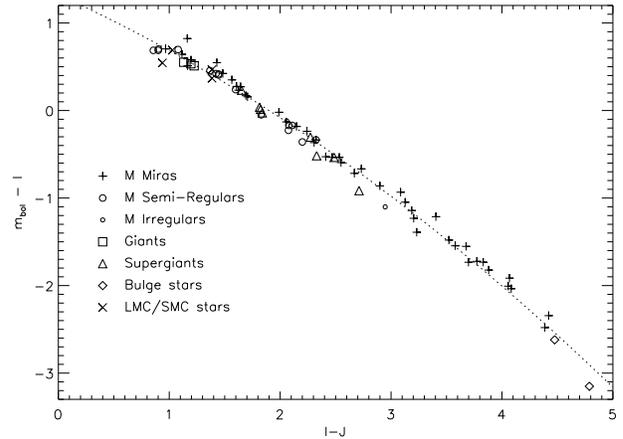}}
  \caption[]{The bolometric correction BC$_{\rm I}$ as a function of the 
  I$-$J colour for cool oxygen-rich stars. The dotted line is the 
  polynomial fit}
  \label{bc}
\end{figure}

The large spectral region covered by the observations and the fact that 
the cool stars emit most of their energy in the near-infrared enable us to
compute bolometric corrections.
A first order correction to the determination of the bolometric luminosity 
has been applied by extrapolating the observed energy distributions on the 
blue and red sides by synthetic spectra as explained in Sect.~3.3.2. The 
mean correction on the bolometric magnitudes amounts to 0.22~mag.
No additional correction was added in the case of dust shell enshrouded 
stars (see Sect.~5). The zero-point of the bolometric scale is computed with 
the following values (Bessell et al.\ 1998): 
M$_{\rm bol,\odot}$=4.74, M$_{\rm V,\odot}$=4.81, V$_{\odot}$=$-$26.76, 
f$_{\rm tot,\odot}$=1.371$\times 10^6$ erg\,cm$^{-2}$\,s$^{-1}$.
The absolute flux at 5556~\AA\ of the Vega model is set to 3.44\,10$^{-9}$ 
erg\,cm$^{-2}$\,s$^{-1}$~\AA$^{-1}$ (Hayes 1985) to achieve a V magnitude 
of 0.03~mag for Vega (Bessell et al.\ 1998).
Figure~\ref{bc} presents the bolometric correction 
BC$_{\rm I}=$ m$_{\rm bol}-$I versus I$-$J. This relation is of 
particular interest for the ongoing infrared sky surveys like DENIS 
(Epchtein 1998). It is remarkable that local, Bulge and LMC/SMC 
stars, and all variability types follow the very same relation: there is
a single narrow sequence for oxygen-rich stars.
A polynomial fit gives:
\begin{equation}
 \mathrm{m}_\mathrm{bol} - \mathrm{I} = 1.32 - 0.574 \times
 \mathrm{(I-J)} - 0.0646 \times \mathrm{(I-J)}^2
 \label{fitbc}
\end{equation}
The standard deviation of the relation is 0.09~mag.

A complete study of the bolometric corrections derived from this 
sample will be presented elsewhere (Mouh\-cine et al., in preparation).

\section{Extinction}

The colour indices used in this paper have not been corrected for 
extinction. To what extent may interstellar and circumstellar matter 
bias the two-colour relations and add to their dispersion?

In order to estimate the effects of interstellar extinction on the 
colours, we used the extended mean extinction law given by Fluks et al.\ 
(1994) based on the observed mean extinction law of Savage \& Mathis (1979) 
and on the theoretical extinction law of Steenman \& Th\'e (1989, 1991).
From the tabulated $R(\lambda) = a(\lambda)/E_{B-V}$ values, we compute 
the interstellar extinction correction for band $B_{i}$ as a function of 
$E_{B-V}$:
\begin{equation}
  B_{i} - B_{i,0} = - 2.5 \log \frac{\int_{0}^{\infty} S_{i}(\lambda) 
  F(\lambda) d\lambda}{\int_{0}^{\infty} S_{i}(\lambda) F(\lambda) 
  10^{0.4 R(\lambda) E_{B-V}} d\lambda}
  \label{extinc}
\end{equation}
where $B_{i}$ is the uncorrected index, $B_{i,0}$ is the dereddened index, 
$S_{i}(\lambda)$ is the response function within the band $B_i$ and
$F(\lambda)$ is the observed spectrum. As an approximation, we take 
$F(\lambda)$ as the arithmetic mean of all the spectra.

\begin{figure}
  \resizebox{\hsize}{!}{\includegraphics{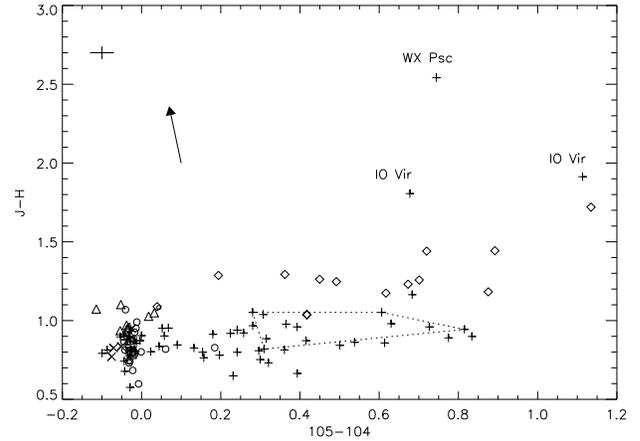}}
  \caption[]{J$-$H versus the near-infrared \Teff\ indicator 105$-$104. 
  The arrow gives the extinction vector for an $E_{B-V}$ equal to 1. The 
  cross in the upper left corner gives the typical error bars. Lines 
  connect observations of X\,Men, an M3 Mira, at various phases. Same 
  symbols as Fig.~\ref{ti}}
  \label{jh}
\end{figure}

To first order, reddening mimics a shift towards lower temperatures in
all the usual broad band two-colour plots (e.g.\ Fig.~\ref{ti+}). But 
molecular band indices may be little affected or even vary orthogonally 
to the temperature sequence, depending on the separation between the two 
photometric passbands used, and on the side of the molecular absorption 
feature on which the ``continuum'' is measured. Thus, the effect of dust 
is expected to be most obvious in plots that combine a broad band colour 
and a molecular index. An example is given in Fig.~\ref{jh}.
Most of the local Miras are grouped along a sequence at J$-$H $\simeq$ 
0.85. The dispersion across the sequence is real, i.e.\ much larger than
observational errors. But only a small fraction of the dispersion may be
associated with interstellar extinction: the motion of individual
stars in J$-$H during their pulsation cycle is comparable with the width
of the sequence (see the example of the M3 Mira X\,Men on Fig.~\ref{jh}), 
while the effect of typical values of the optical depth towards local 
stars is much smaller (comparable in amplitude to the observational 
uncertainties). This justifies neglecting interstellar extinction 
towards local stars.

\begin{figure}
  \resizebox{\hsize}{!}{\includegraphics{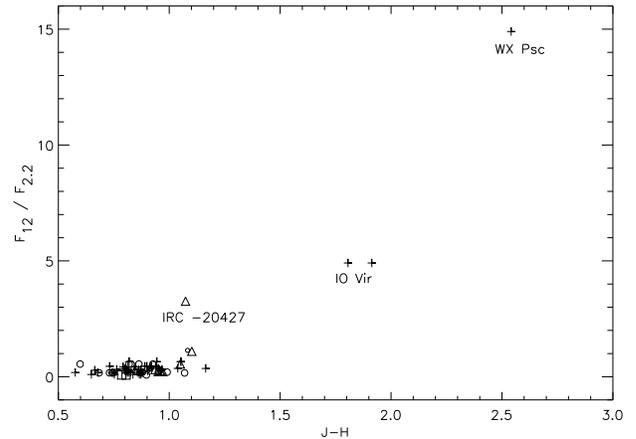}}
  \caption[]{The F(12~$\mu$m)/F(2.2~$\mu$m) ratio versus J$-$H. 
  Three stars (WX\,Psc, IO\,Vir and IRC\,$-$20427) present an infrared
  excess due to the presence of a cool circumstellar shell. Same 
  symbols as Fig.~\ref{ti}}
  \label{ie}
\end{figure}

Two local Miras with very large J$-$H values (larger than 1.5) are
identified in Fig.~\ref{jh}. IO\,Vir, a M6 Mira, was observed twice. 
It is a known OH maser (OH\,334.7+50.0, Turner 1979). This star is thus 
enshrouded in an important dust shell. So is the second extreme Mira, 
WX\,Psc (OH\,128.6-50.1). IO\,Vir and WX\,Psc are the only two Miras of 
our sample with an OH designation: this corroborates the fact that 
circumstellar matter, as opposed to interstellar matter, is responsible 
for their red colours.
Infrared excess is measured by the ratio F(12~$\mu$m)/F(2.2~$\mu$m)
(see Habing 1996 and references therein; Le Bertre \& Winters 1998).
This ratio has been derived from the IRAS PSC (IRAS Science Team 1988)
and from the K magnitude (Gezari et al.\ 1996), whenever available (using 
the calibration of Beckwith et al.\ (1976) to convert K magnitudes into 
fluxes at 2.2~$\mu$m). As expected, Fig.~\ref{ie} clearly shows that 
WX\,Psc and IO\,Vir (and to some extent IRC\,$-$20427) have large 
F(12~$\mu$m)/F(2.2~$\mu$m) ratio. The presence of cool circumstellar 
material around these stars is thus confirmed and explains their large 
J$-$H colours.

Even in the extremely obscured cases mentioned above, the data do not 
indicate a significant contribution of circumstellar continuum emission at 
wavelengths shorter than 2.4~$\mu$m. Such a warm dust contribution would 
reduce the equivalent widths of the CO bands (Tanaka et al.\ 1996). No 
significant reduction is observed in our essentially O-rich sample. The 
depths and shapes of the CO bands ($\lambda \geq$ 2.29~$\mu$m) of the 
naturally reddened spectra compare very well to those of cool unobscured 
Miras like X\,Men, once the latter have been artificially reddened to match
the colours of the former. 

The Bulge stars systematically lie above the sequence of local optical
LPVs. The interstellar extinction effect is obvious. The extinction 
towards the Sgr\,I and NGC\,6522 galactic Bulge fields, to which the 
Bulge stars of our sample belong, corresponds to $E_{B-V} \simeq 0.55$ 
according to estimates in the literature (Wood \& Bessell 1983; Glass et 
al.\ 1995; Barbuy et al.\ 1998). Once corrected for reddening, the 
observed Bulge stars lie on the edge of the local Mira sequence. This is 
consistent with the fact that, for Bulge stars, a large J$-$K index has 
been used as a selection criterion for inclusion in the spectroscopic
sample. However, the spectra indicate that at least some of the observed 
Bulge stars are reddened more than the average Sgr\,I and NGC\,6522 field 
stars: for instance, the spectrum of Sgr\,I N$^\circ$11 (observed in July 
1996) is very well matched, in terms of the energy distribution, the 
molecular bands and the global aspect of metal line blends, by the 
spectrum of the local Mira RS\,Hya (February 1995) if $E_{B-V} = 1$. An 
equally good match is obtained between the local X\,Men (March 1996) and 
NGC\,6522 N$^\circ$435 (May 1996) with $E_{B-V} = 1.6$. This additional 
extinction may be due in various proportions to circumstellar matter and
to the patchy distribution of interstellar dust. Similarly high extinction 
values are found towards various star clusters in the neighbourhood of the 
Galactic Centre (Barbuy et al.\ 1998). Only two of the Bulge stars of the
sample have been detected by IRAS (NGC\,6522 N$^\circ$435= IRAS\,17593-3006,
NGC\,6522 N$^\circ$205=IRAS 17591-2959; Glass 1986; Glass et al.\ 1995).
Using the K magnitudes of Wood \& Bessell (1983), we obtain 12~$\mu$m to
2.2~$\mu$m flux ratios of 1.05 and 1.6, respectively, i.e.\ relatively
low values as compared to the local IRAS sources of Fig.~\ref{ie}.
Again, the strong CO absorption longward of 2.29~$\mu$m argues against a 
significant contribution of warm circumstellar dust emission to the 
near-IR light of the Bulge LPVs. 

The Bulge stars tend to display strong VO and TiO bands; molecular bands 
need to be observed in  a larger unbiased sample in order to determine 
whether this is or is not due to the selection of red and luminous objects.

\section{Water in Long Period Variable stars}

The most distinctive features in the near-IR spectra of pulsating red 
giants are their impressive H$_2$O absorption bands. Johnson \& M\'endez 
(1970) already observed Mira type LPVs with much deeper water vapour 
bands than static giants of similar spectral types. LPV models (Bessell et 
al.\ 1989b; Bessell et al.\ 1996) and observations (e.g.\ Matsuura et 
al.\ 1999) have subsequently confirmed that deep H$_2$O features are related 
to pulsation and have provided basic understanding: pulsation produces much 
more extended atmospheres, and in addition dense cool layers may result 
from the periodical outwards running shocks. 
In various ways pulsation thus leads to the existence of regions where 
relatively low temperatures ($\sim$ $10^3$~K or lower) are combined with 
relatively high densities, conditions that favour the formation of H$_2$O 
and possibly even of dust (Bowen 1988; Feuchtinger et al.\ 1993; Jeong et 
al.\ 1999).

Figure~\ref{ti+} (top left) shows that H$_2$O absorption indices for LPVs
range from 0 to values far above those the coolest static giants reach. At 
a given stellar energy distribution a large range of H$_2$O absorption 
indices is found (a similar phenomenon has been described for SiO bands by
Aringer et al.\ 1999). Even relatively warm LPV spectra (instantaneous 
\Teff\ $\geq$ 3500~K) may display deep water vapour bands, a behaviour that 
remains to be reproduced quantitatively by LPV models (Mouhcine \& 
Lan\c{c}on 1998). More data (in particular complete phase coverage for more 
stars) is required to gain better insight into the relation between the 
H$_2$O features, pulsation amplitude, phase, metallicity and details of the 
stellar structure.

For the purposes of this paper, we conclude that deep H$_2$O features can 
be taken as a clear indication for pulsation, while the absence of water 
vapour does not exclude that the observed star be an LPV.

\section{Supergiants, giants and variability types}

From a spectrophotometric point of view, supergiants, giants and dwarfs 
are usually separated mainly on the basis of CO and H$_2$O absorption, 
which respectively strengthens and weakens from luminosity class V to I
(e.g. Aaronson et al.\ 1978; Kleinmann \& Hall 1986; Origlia et al.\ 
1993). The strength of various metal lines, e.g.\ the Na doublet at 
2.20~$\mu$m and the Ca triplet at 2.26~$\mu$m, also depend on the 
luminosity class. Ram\'{\i}rez et al.\ (1997) found that the quantity 
log[EW(CO)/(EW(Na) + EW(Ca))] is an excellent luminosity indicator,
at least for K and M giants and dwarfs. Bessell et al.\ (1989a) identified 
the CN band at 1.1~$\mu$m as a surface gravity sensitive feature.

\begin{figure}
  \resizebox{\hsize}{!}{\includegraphics{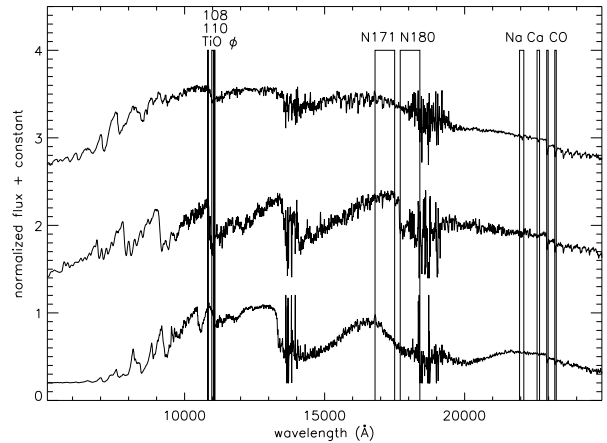}}
  \caption[]{Filters used in Sect.~7. Same spectra as Fig.~\ref{vf1}}
  \label{vf2}
\end{figure}

In this section, we summarise the results of a systematic search for 
indices that separate the giants from supergiants of our sample. 
Our sample contains 10 supergiant observations: 3 non-variable supergiants,
plus 3 Semi-Regular variables of the SRc type, plus 3 Irregular variables 
of the Lc type, one of them observed twice. All these supergiants are of
spectral type M0 or later. The sample stars being cooler than those of 
Ram\'{\i}rez et al.\ (1997) and our spectral resolution lower, the CaI and 
NaI features around 2.2~$\mu$m are seriously blended with molecular 
absorption features, leading us to redefine Na and Ca indices. 
The passbands and indices used are defined in Table~\ref{filt1}.
The passbands for CO measurements around 2.3~$\mu$m are taken from
Kleinmann \& Hall (1986). F180W and F171W correspond to the Hubble Space 
Telescope narrow band NICMOS filters. They were initially designed to 
measure the strength of the 1.7~$\mu$m C$_2$ absorption band in carbon 
stars (Thompson, private communication), but actually provide a good H$_2$O 
absorption index (see Lan\c{c}on et al.\ 1999 for applications 
to stellar population studies). Other common H$_2$O indices (e.g.\ based on
Wing filters at 2.0 and 2.2~$\mu$m) would lead to qualitatively similar 
results. The selected filters are displayed on Fig.~\ref{vf2}.

\begin{table}
\caption[]{Filters used to define indices liable to separate giants and 
Miras from supergiants}
\begin{flushleft}
\begin{tabular}{llll}
\hline\noalign{\smallskip}
Filter & Center & Width & Comments \\
name   & (\AA)  & (\AA) &          \\
\noalign{\smallskip}
\hline\noalign{\smallskip}
F171W & 17150 & 700 & HST NICMOS band \\
F180W & 18050 & 700 & HST NICMOS band, H$_2$O\,+\,C$_2$ \\
xNac$_1$ & 21920 & 20 & 'continuum' \# 1 for Na \\
xNac$_2$ & 21960 & 20 & 'continuum' \# 2 for Na \\
xNa & 22036 & 152 & Na I doublet \\
xNac$_3$ & 22120 & 20 & 'continuum' \# 3 for Na \\
xNac$_4$ & 22168 & 25 & 'continuum' \# 4 for Na \\
xCac$_1$ & 22548 & 75 & 'continuum' \# 1 for Ca \\
xCa & 22630 & 90 & Ca I triplet \\
xCac$_2$ & 22703 & 55 & 'continuum' \# 2 for Ca \\
xCOc & 22899 & 52 &  'continuum' for CO \\
xCO$_1$ & 22957 & 52 & $^{12}$CO(2,0) band-head \\
xCO$_2$ & 23245 & 54 & $^{12}$CO(3,1) band-head \\
\noalign{\smallskip}
\hline
\noalign{\smallskip}
\multicolumn{4}{l}{
CO = $-2.5\,\log [$(xCO$_1+$xCO$_2)/$xCOc$] \ - $CO(Vega) }\\
\multicolumn{4}{l}{
Na = $-2.5\,\log [$xNa/(xNac$_1+\ldots +$xNac$_4)] \ - $Na(Vega) }\\
\multicolumn{4}{l}{
Ca = $-2.5\,\log [$xCa/(xCac$_1+$xCac$_2)] \ - $Ca(Vega) }\\
\multicolumn{4}{l}{
H$_2$O = $-2.5\,\log [$F180W/F171W$] \ - $H$_2$O(Vega) }\\
\noalign{\smallskip}
\hline
\end{tabular}
\end{flushleft}
\label{filt1}
\end{table}

\begin{figure}
  \resizebox{\hsize}{!}{\includegraphics{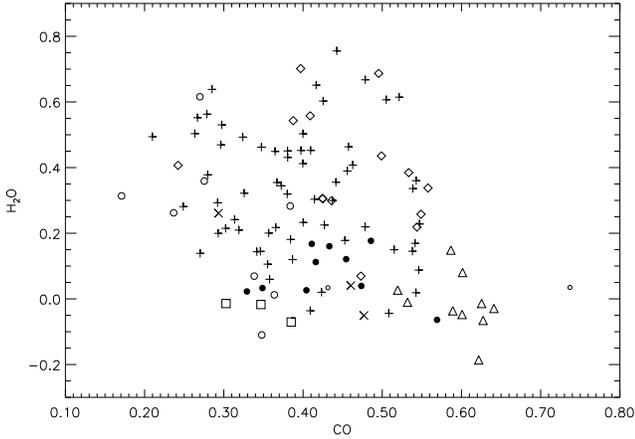}}
  \caption[]{The H$_2$O index versus the CO index (see Sect.~7). Same
   symbols as Fig.~\ref{ti} except for M Semi-Regulars (empty circles: 
   SRa; filled circles: SRb). Supergiants (triangles) are discriminated in 
   this diagram}
  \label{gsg1}
\end{figure}

\begin{figure}
  \resizebox{\hsize}{!}{\includegraphics{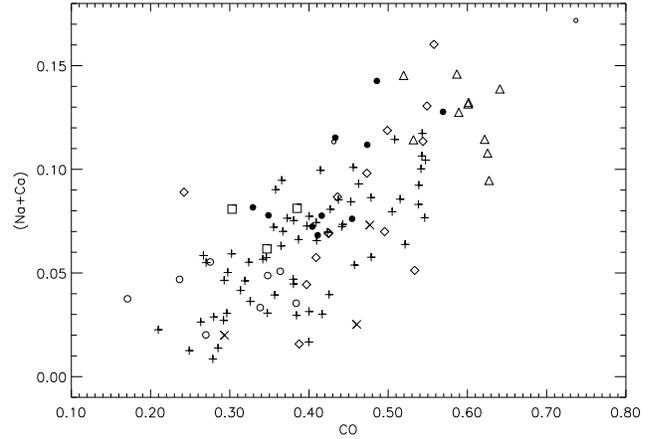}}
  \caption[]{The (Na+Ca) index versus the CO index (see Sect.~7). 
   Same symbols as Fig.~\ref{gsg1}}
  \label{gsg2}
\end{figure}

The CO, H$_2$O and metal line measurements are plotted in 
Fig(s).~\ref{gsg1} and \ref{gsg2}. Supergiants systematically provide the 
highest metal line and CO indices of the sample of O-rich stars (carbon 
stars are discussed in Sect.~8). As expected, their mean CO and metal 
absorption is high and their mean H$_2$O absorption low, compared to the 
mean properties of the M type variable giants of luminosity class III. The 
combined constraints  CO $\geq$ 0.51, H$_2$O $\leq$ 0.15 and (Na+Ca) $\geq$ 
0.09 isolate the supergiants with only 3 intruders: one of the two 
observations of KV\,Car (SRb variable), one of the three observations of 
WW\,Sco (Mira variable) and the observation of V774\,Sgr (Lb variable).

The luminosity class sensitivity of the CO/(Na+Ca) combination of 
Ram\'{\i}rez et al. (1997) indeed extends to giant and supergiant 
luminosities. However, since (at a resolution of 1100) the centers of the 
Na and Ca line blends lie typically 10\,\% below the so-called continuum 
and their range of variation is small compared to the range of variation 
of CO absorption, the combined index carries information similar to the CO
index itself.

\begin{figure}
  \resizebox{\hsize}{!}{\includegraphics{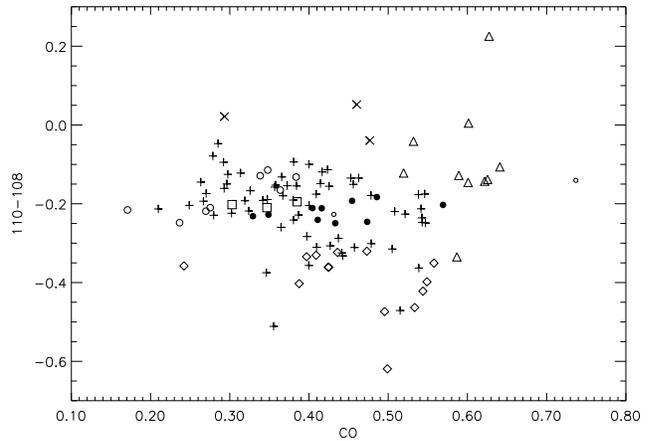}}
  \caption[]{The CN absorption index 110$-$108 versus the CO index.
   Same symbols as Fig.~\ref{gsg1}}
  \label{gsg3}
\end{figure}

\begin{figure}
  \resizebox{\hsize}{!}{\includegraphics{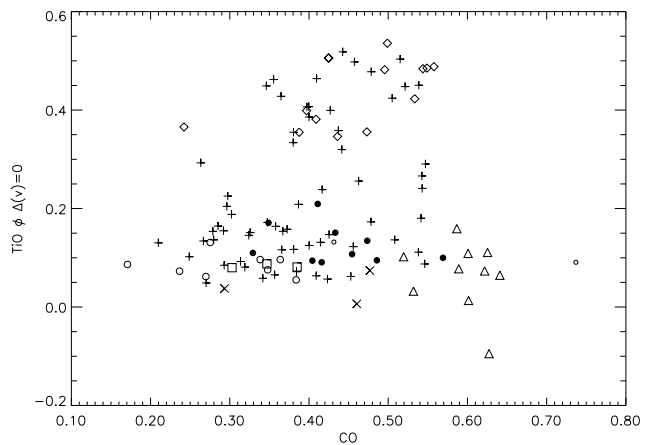}}
  \caption[]{The absorption index of TiO $\phi$ ($\Delta$v=0) versus 
   the CO index. Same symbols as Fig.~\ref{gsg1}}
  \label{gsg4}
\end{figure}

The isolation of the supergiants appears particularly clearly when a CO 
measurement is combined with the gravity indicator of Bessell et al.\ 
(1989a), the CN index 110-108 (Fig.~\ref{gsg3}).
In the same way, TiO indices measuring the $\phi$ bands at 1.1 or 
1.2~$\mu$m are absent in supergiants, and provide an alternative 
segregation tool (Fig.~\ref{gsg4}). These TiO indices are useful to
identify very cool AGB stars.
The 108, 110 and TiO $\phi$ filters are shown in Fig.~\ref{vf2}.

Figures \ref{gsg1}--\ref{gsg4} also indicate a tendency for SRa type LPVs 
to have a lower CO index than SRb type variables. According to Kerschbaum 
\& Hron (1992, 1994), SRa type variables appear as intermediate objects 
between Miras and SRb variables in several aspects. This assertion seems 
supported by Fig(s).~\ref{gsg1}--\ref{gsg4} concerning the CO absorption
band. We note nevertheless that only 2 SRa variables (adding up to 8 
observations) and 4 SRb variables (10 observations) have been observed. 
More spectroscopic data should be studied to confirm this effect.

\section{Chemical types}

The spectra of carbon stars have little in common with those of oxygen-rich 
giants appart from CO absorption and a red energy distribution 
(Fig.~\ref{vf2}). While, according to both observations and model 
atmospheres, it might be possible to identify selected regions of the 
near-IR spectra of O-rich LPVs as quasi-continuum windows (e.g.\ 
1.04~$\mu$m, 2.15~$\mu$m), the whole optical and near-IR range of C-rich 
stars is shaped by molecular bands of carbon based molecules. In the 
wavelength range under consideration the C-rich spectra display much less 
temperature and pulsation phase dependence than O-rich spectra.

Our sample includes 26 NIR spectra of 6 carbon-rich stars. Four are 
Semi-Regular LPVs (7 spectra of RU\,Pup, 6 of Y\,Hya, 3 of T\,Cae, 1 of 
S\,Cen). The two Mira-type variables are the dust-enshrouded R\,Lep and 
the SC star BH\,Cru. Although the sample may not be representative of all 
C-rich stars, it allows us to demonstrate which spectrophotometric indices 
are useful in the separation of chemical subtypes.

It is well known that stars can be to a certain point separated into carbon
stars and oxygen stars using broad band colour--colour diagrams (e.g.\
Frogel et al.\ 1980, 1990; Feast 1994; Zijlstra et al.\ 1996; Loup et al.\ 
1998). C-rich stars appear in general redder than O-rich stars in such 
diagrams, and as a consequence, an often heard statement is that carbon 
stars are the reddest giants. 

The present data shows that this is not in general true. It is based on 
Magellanic Cloud data and is likely to hold for metal-deficient 
environments, where few if any very cool O-rich AGB stars are expected to 
exist because dredge-up of carbon produces C-stars more efficiently, but it 
must be reconsidered in the solar neighbourhood or in more general surveys. 
Excluding objects of our sample for which the presence of a circumstellar 
envelope is known, the O-rich LPV sequence reaches much redder values than 
the C-rich sequence in most colours (e.g.\ V$-$I, V$-$K, R$-$K, I$-$K). The 
colour distribution is similar for both chemical types in H$-$K. 

\begin{figure}
  \resizebox{\hsize}{!}{\includegraphics{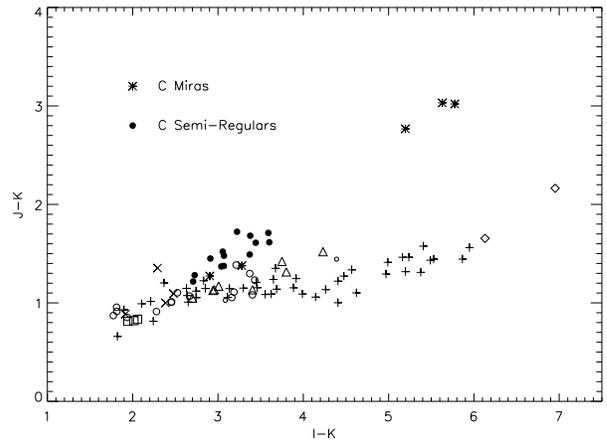}}
  \caption[]{J$-$K versus I$-$K. Carbon-rich Miras are represented by 
  asterisks and carbon-rich Semi-Regulars by filled circles. Otherwise,
  same symbols as Fig.~\ref{ti}}
  \label{ijk}
\end{figure}

Most pure optical/NIR colours are thus insufficient to enable an unambiguous 
discrimination if used alone, and colours strongly affected by the specific 
molecular features of carbon stars are needed. Zijlstra et al.\ (1996) 
suggest the infrared colour K$-$[12], in which C-stars appear relatively red 
due to HCN and possibly C$_2$H$_2$ absorption in the 12--14$\mu$m range 
(Hron et al.\ 1998).  J$-$K also is a useful indicator. Due to CN and C$_2$
absorption in the J band, the C-stars appear systematically redder in this 
colour than O-rich stars with, for instance, identical I$-$K values
(Fig.~\ref{ijk}).
J$-$K $\geq$ 1.5 or 1.6 has been used for the separation of C-stars in LMC
surveys (Loup et al.\ 1998; Wood et al.\ 1999). O-rich spectra with higher 
J$-$K values exist among the stars with no obvious dust shells of our 
sample, but they are rare. They generally correspond to Mira-type variables 
observed at minimum light. The J$-$K criterion remains applicable at solar 
metallicities, but it misses the warmer of the C-stars and only holds if 
M-stars with strong mass loss (and circumstellar emission) have previously 
been excluded. The narrow-band molecular indices discussed below avoid 
confusion.

\begin{figure}
  \resizebox{\hsize}{!}{\includegraphics{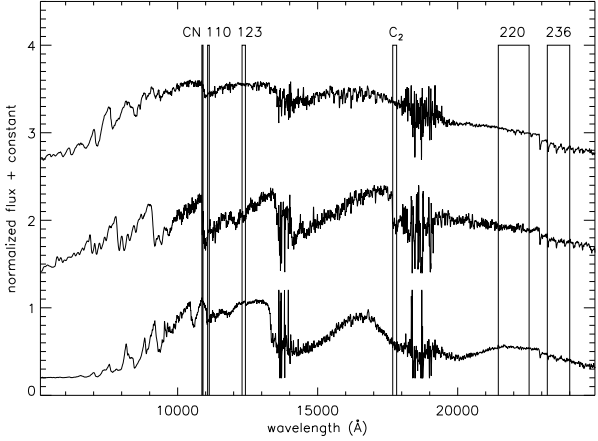}}
  \caption[]{Filters used in Sect.~8. Same spectra as Fig.~\ref{vf1}}
  \label{vf3}
\end{figure}

The most obvious features identifying C-star spectra are those of CN and 
C$_2$ bands. Indices for CN measurements include the Wing pairs 110$-$108 
or 110$-$123, as well as the more selective narrow band index based on the 
first 2 filters in Table~\ref{filt2}. Indices sensitive to the steep 
Ballik-Ramsay C$_2$ bandhead at 1.77~$\mu$m (Ballik \& Ramsay 1963;
Goorvitch 1990) include the HST NICMOS intermediate band index 
F180W$-$F171W and the narrow band index based on the nearly adjacent last 
2 filters in Table~\ref{filt2}. The filters are shown in Fig.~\ref{vf3}.

\begin{table}
\caption[]{Filters used to define indices which measure carbon species}
\begin{flushleft}
\begin{tabular}{llll}
\hline\noalign{\smallskip}
Filter & Center & Width & Comments \\
name   & (\AA)  & (\AA) &          \\
\noalign{\smallskip}
\hline\noalign{\smallskip}
xCN$_{\rm cont}$ & 10840 & 30 & 'continuum' for CN \\
xCN & 10885 & 40 & CN A--X $\Delta$v=0 transition \\
xC$_{2,\rm cont}$ & 17570 & 100 & 'continuum' for C$_2$ \\
xC$_2$ & 17750 & 140 & Ballik-Ramsay band \\
\noalign{\smallskip}
\hline
\end{tabular}
\end{flushleft}
\label{filt2}
\end{table}

\begin{figure}
  \resizebox{\hsize}{!}{\includegraphics{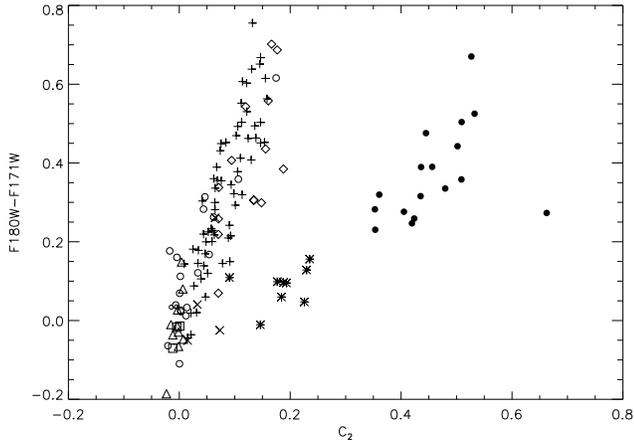}}
  \caption[]{The NICMOS index F180W$-$F171W versus the C$_2$ index. Same 
  symbols as Fig.~\ref{ti} and Fig.~\ref{ijk}}
  \label{ec2}
\end{figure}

In Fig.~\ref{ec2}, the NICMOS index F180W$-$F171W is plotted against the 
narrow band C$_2$ index. The good correlation among each sequence comes 
from the overlap between the passbands used for both indices. However, only 
in the narrow band index are the two filters close enough not to be 
sensitive to the broad absorption band of H$_2$O. While the NICMOS colour 
is able to identify LPVs independently of their chemical type, it is not 
capable of separating C-rich objects from O-rich ones (Lan\c{c}on et al.\ 
1999).

\begin{figure}
  \resizebox{\hsize}{!}{\includegraphics{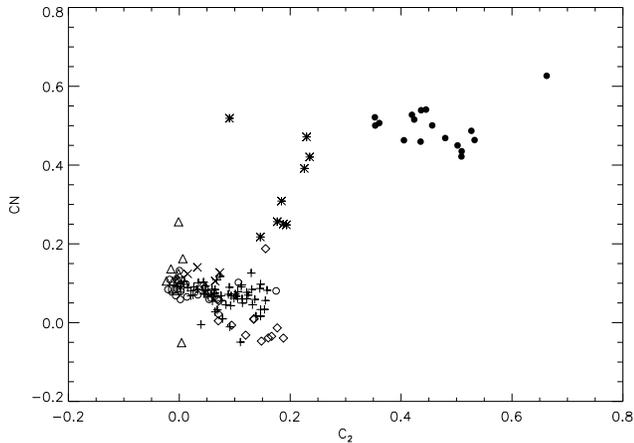}}
  \caption[]{The CN index versus the C$_2$ index. Same symbols as 
  Fig.~\ref{ti} and Fig.~\ref{ijk}}
  \label{cnc2}
\end{figure}

Figure~\ref{cnc2} shows the CN index against the C$_2$ index. Once again, 
the oxygen-rich stars are clearly separated from carbon-rich stars as 
expected. The O-rich stars are now grouped together instead of forming 
a sequence.

Another noticeable feature in Fig(s).~\ref{ec2} and \ref{cnc2} is the 
apparent segregation between carbon Miras and carbon Semi-Regular 
variables. If, on average, Mira-type pulsation is reached at a later
evolutionary stage on the AGB than Semi-Regular pulsation, as suggested 
by recent studies of period--luminosity sequences in the LMC (Wood et al.\ 
1999), and if the C$_2$ band strength depends primarily on the C/O ratio, 
which on average increases with time along the AGB, then one expects to find 
the C-rich Miras at higher C$_2$ indices than the Semi-Regulars. This
simple global scenario cannot be applied to samples as small as this 
one. In particular, the SC star BH\,Cru pulsates like a Mira variable 
although its C/O ratio is close to 1. Larger and more representative samples 
are required to study systematic effects of the pulsation type.

The CO indices measured for carbon stars cover the full range observed
for O-rich objects, including supergiant stars. In particular, the 
2.3~$\mu$m CO absorption of the SC star BH\,Cru compares to the strongest 
band observed in our supergiant sample. This must be taken as a warning 
against the exclusive use of the CO index in searches for red supergiant 
stars (although the rarity of SC type stars prevents this finding from 
being of major concern in the study of integrated spectra of stellar 
populations).

\begin{figure}
  \resizebox{\hsize}{!}{\includegraphics{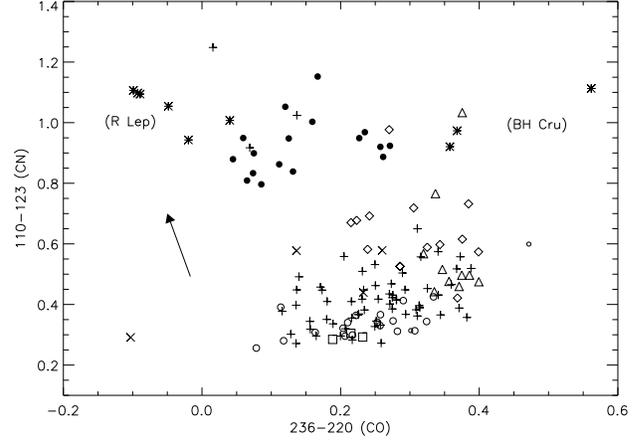}}
  \caption[]{The 110$-$123 index versus the 236$-$220 index. Same 
  symbols as Fig.~\ref{ti} and Fig.~\ref{ijk}. The six observations 
  of the C-Mira R\,Lep and the three observations of the SC-Mira BH\,Cru 
  are indicated. The arrow gives the extinction vector for an $E_{B-V}$ 
  equal to 1}
  \label{cw}
\end{figure}

As already mentioned in the case of C$_2$, the comparison of the CO index 
defined in Sect.~7 with the larger baseline index 236$-$220 illustrates 
the advantages of using narrow filters located very close to the feature 
of interest (Fig.~\ref{cw}): the extremely low 236$-$220 index of R\,Lep 
is merely due to its extremely red energy distribution; the systematic 
shift in CO between the 236$-$220 values of C-rich and O-rich stars is 
explained by additional molecular absorption in the 220 band; the shift in 
both CO and CN between Bulge and solar neighbourhood objects is mainly the 
effect of extinction. Colour corrections are required to circumvent 
ambiguities in the large baseline indices (Frogel \& Whitford 1987).

\section{\cc\ ratio}

The \cc\ ratio in red giants is an important indicator of nuclear 
processing that occurs inside a star. The first dredge-up occurs when the 
star ascends the first giant branch: the number ratio \cc\ drops from its
initial value of $\sim$ 90 to between 18 and 26 (Charbonnel 1994).
The intermediate-mass stars ($\geq$ 3--4~$\cal{M}_\odot$) undergo a second 
dredge-up at the beginning of the AGB: the \cc\ ratio decreases a little
bit more for these stars.
Then, the \cc\ ratio increases along the Thermally-Pulsating Asymptotic 
Giant Branch as the third dredge-up enriches the stellar surface with 
freshly synthesized elements (e.g.\ Forestini \& Charbonnel 1997).
However, when Hot Bottom Burning occurs, this ratio drops again to small
values (Boothroyd et al.\ 1993). The \cc\ isotope ratio is thus likely to
exhibit a large spread of values for a given sample of cool giants.
Our set of spectra enables us to check for an evolution of the \cc\ ratio.

\begin{figure}
  \resizebox{\hsize}{!}{\includegraphics{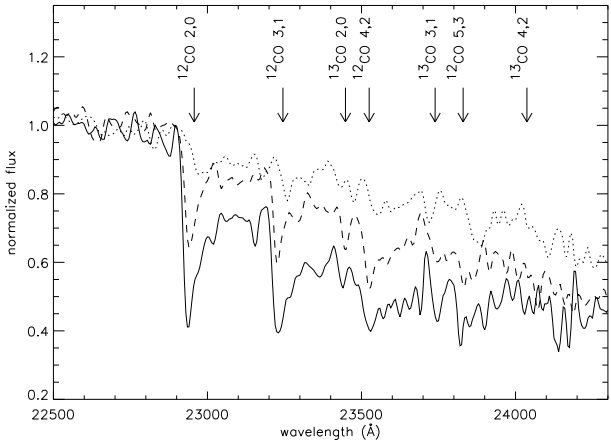}}
  \caption[]{Spectra exhibiting \cod\ and \cot\ lines with decreasing 
  strength (solid line: BH\,Cru, an SC Mira; dashed line: RS\,Hya, an M Mira; 
  dotted line: S\,Car, an M Mira)}
  \label{co}
\end{figure}

Figure~\ref{co} shows three examples of star which exhibit \cod\ and 
\cot\ lines with decreasing strength (solid line: SC4.5--SC7 Mira BH\,Cru;
dashed line: M6 Mira RS\,Hya; dotted line: K5--M6 Mira S\,Car). 
Positions of the CO bandheads are shown. The spectra are normalized to a 
common flux (F$_{\lambda}$) around 22\,900~\AA.

\begin{figure}
  \resizebox{\hsize}{!}{\includegraphics{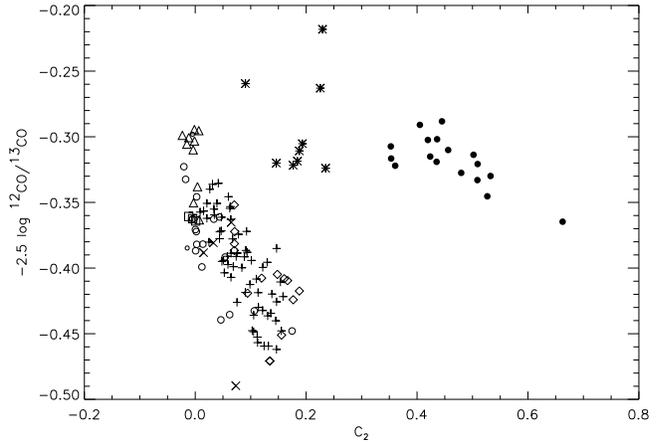}}
  \caption[]{The \coco\ index as a function of the C$_2$ index. Same 
  symbols as Fig.~\ref{ti} and Fig.~\ref{ijk}}
  \label{coc2}
\end{figure}

A semi-quantitative approach using photometric indices is illustrated in 
Fig.~\ref{coc2}. A \coco\ index is computed from the fluxes in  rectangular 
80\,\AA\ wide filters, respectively centered in the four \cod\ and the three 
\cot\ bandheads shown in Fig.~\ref{co}. It is expected to vary with the \cc\ 
ratio. It is plotted in Fig.~\ref{coc2} against the C$_2$ index 
(Table~\ref{filt2}) which separates C-rich from O-rich stars. Interstellar 
extinction has a negligible effect on the indices of Fig.~\ref{coc2}.

The C-rich stars exhibit the larger \coco\ indices: as expected, the \cc\ 
ratio increases as the star undergoes third dredge-up that turns it into a 
carbon star.
Supergiants (triangles) have also large \coco\ indices compared to giants
and LPVs. It may not be due to intrinsic large values of \cc. Indeed, 
predicted \cc\ ratios from computations are weakly dependent on stellar 
mass at $\cal{M} \geq$ 2~$\cal{M}_\odot$ (El Eid 1994): supergiants are
not expected to exhibit large \cc\ ratios. On the contrary, supergiants have 
in most cases low observed \cc\ ratios ($\alpha$\,Ori and $\alpha$\,Sco for 
instance, Harris \& Lambert 1984). More probably, the large \coco\ are due 
to radiative transfer effects on the CO bands (luminosity, atmospheric 
extension, line blends with other species).

Oxygen-rich LPVs exhibit a large spread of \coco\ indices, sometimes
with values comparable to C-rich stars. 
Model atmospheres would be necessary to carefully determine the \cc\ ratios 
of our sample stars and explain some of the trends in Fig.~\ref{coc2},
and compare to predicted computations from evolutionary models.

\section{Conclusion}

We have explored the large sample of optical and near-infrared spectra
of Lan\c{c}on \& Wood (1998 and in preparation), and derived useful tools to 
connect spectrophotometric signatures with fundamental properties of cool 
oxygen- and carbon-rich LPVs and supergiants.

Temperatures of oxygen-rich stars are assigned on the basis of synthetic 
spectra, computed with up-to-date oxygen-rich model atmosphere grids 
(Plez 1992; Plez, private communication) and adjusted to the optical data. 
They range from 2500~K to 4000~K. It is recalled that the relationship 
between this semi-empirical temperature and effective temperature is 
expected to depend on pulsation properties as well as on the more usual 
physical attributes of a star (metallicity, surface gravity). 
While several optical colours appear to be rather good indicators of the 
assigned temperature (e.g.\ 710$-$888), only a few purely near-infrared 
temperature indicators were found. The Wing colour 105$-$104 and the 
broad-band colours I$-$J and I$-$K can be used as a first approximation. 
The lack of better near-infrared temperature indicators is unfortunate for 
extragalactic survey purposes. 
V$-$K, although not a purely NIR colour (and thus also more sensitive
to extinction), proves to be a very convenient indicator of the optically
assigned temperature over the whole range exhibited by the sample stars.

The large spectral region covered by the observations enables us to derive
a relation between the bolometric correction BC$_{\rm I}$ and the 
broad-band colour I$-$J for oxygen-rich cool stars. The standard deviation 
is only of 0.09\,mag. It can be applied as a first approximation to red 
giants, supergiants and LPVs. This relation is of particular interest for 
ongoing infrared sky surveys.

Interstellar extinction towards most stars of the solar neighbourhood
sample appears to be small. 
Only four local stars (IO\,Vir, WX\,Psc, R\,Lep and IRC\,$-$20427) are 
clearly dust enshrouded. Even the latter display deep CO bands at 
2.3~$\mu$m, arguing against a significant contamination by circumstellar 
dust emission in the near-IR.

Deep H$_2$O features present in the near-IR spectra are a strong indication 
of pulsation. Nevertheless, the absence of water vapour does not exclude 
that the observed star is an LPV.

Luminosity class I stars can be efficiently distinguished from luminosity
class III stars with appropriate combinations of indices based on CO, 
H$_2$O, Na, Ca, CN or TiO lines. SRa type LPVs show a tendency for lower 
CO indices than SRb variables.

In order to separate carbon-rich LPVs from oxygen-rich stars, colours 
strongly affected by the specific molecular features of carbon stars are 
needed. Selected broad-band colours (e.g.\ J$-$K) are useful indicators, 
but narrow-band molecular indices based on CN and C$_2$ bands are by far 
more efficient.

A \coco\ index has been computed from the empirical spectra. As expected
if the \coco\ index actually varies with the \cc\ ratio, the C-rich stars 
exhibit on average larger index values than the O-rich stars. The 
supergiants of the sample also exhibit large \coco\ indices. This behaviour 
is opposite to what is expected from evolution models. This might be
due to radiative transfer effects on the CO bands (luminosity, atmospheric 
extension, line blends with other species).

The different indices and relations discussed are useful in the 
interpretation of integrated NIR spectra of stellar populations and 
of the considerable amount of data provided by ongoing infrared sky surveys, 
as well as in infering stellar properties of individual cool giants and 
supergiants. However, in order to resolve remaining ambiguities and
to gain quantitative information, the dependence of the observable indices 
on stellar parameters, surface abundances, environement, etc, must be more 
extensively investigated. Only comparisons to models will eventually permit 
to complete this work. Extreme difficulties result from the fact that 
many complex aspects of LPV modelling (oxygen-rich/carbon-rich 
transition, pulsational distinction between Mira and Semi-Regular variables, 
effects of pulsation on the atmosphere, non-grey emergent spectra, etc.) 
should be taken into account simultaneously. Some recent and important 
progress allows us to hope that this formidable task might be carried out
in a future not completely out of reach.

\begin{acknowledgements}
RA gratefully thanks D.\ Egret and J-L.\ Halbwachs for their hospitality 
and for providing financial support while he was staying at the Strasbourg 
Observatory. RA benefits of a EU TMR ``Marie Curie'' Fellowship at ULB.
Part of this work was completed while BP was at Uppsala Astronomiska 
observatoriet and at the Atomspektroskopi division in Lund. B.\ Gustafsson 
and S.\ Johansson are both thanked for their support and hospitality.
This research has made use of the Simbad database operated at CDS,
Strasbourg, France.
\end{acknowledgements}

\end{document}